# Behavioral criteria of feedforward processing in rapid-chase theory: Some formal considerations

*arXiv:1405.5795v3 [q-bio.NC]*

Thomas Schmidt, University of Kaiserslautern, Germany
Faculty of Social Sciences, Experimental Psychology Unit
www.sowi.uni-kl.de/psychologie
thomas.schmidt@sowi.uni-kl.de

**Abstract**

*The rapid-chase theory of response priming defines a set of behavioral criteria that indicate feedforward processing of visual stimulus features rather than recurrent processing. These* **feedforward criteria** *are strong predictions from a feedforward model that argues that sequentially presented prime and target stimuli lead to strictly sequential waves of visuomotor processing that can still be traced in continuous motor output; for instance, in pointing movements, muscle forces, or EEG readiness potentials. The feedforward criteria make it possible to evaluate whether some continuous motor output is consistent with feedforward processing, even though the neuronal processes themselves are not readily observable. This paper is intended as an auxiliary resource that states the criteria with some degree of formal precision.*

**1. Introduction**

The classical view of the visual system as a stepwise processing hierarchy neglects temporal aspects, such as the speed of information transfer between areas, the role of re-entrant information, and the role of recurrent processing loops (Bullier, 2004; Gilbert & Sigman, 2007). Recently, researchers have proposed a fundamentally revised view of the visual and visuomotor system, stressing a distinction between two radically different types of visual processing: a rapid feedforward process where visual activation proceeds in bottom-up direction, and a slower, recurrent process developing in the immediate wake of this "*fast feedforward sweep*" (Lamme & Roelfsema, 2000; also see Bullier, 2001; VanRullen & Koch, 2003; VanRullen & Thorpe, 2002). The concept of feedforward vs. recurrent processing now plays a major role in the understanding of fast image categorization (Thorpe, Fize, & Marlot, 1996; VanRullen & Thorpe, 2001), figure-ground segmentation and grouping (Roelfsema, 2006; Houtkamp & Roelfsema, 2010), masked priming (e.g., Schmidt, Haberkamp, Veltkamp, Weber, Seydell-Greenwald, & Schmidt, 2011), and visual awareness (e.g., Lamme, 2010).

The neuronal processes that comprise the feedforward sweep are not readily observed in humans. Therefore, we have argued that some bridging principles are needed that connect the theoretical notion of a feedforward sweep to overt, observable behavior (Schmidt, Niehaus, & Nagel, 2006; Schmidt et al., 2011). We argue that the role of feedforward processing of visual stimuli can be assessed by tracing the time courses of speeded motor responses to sequential stimuli. A paradigm especially suited for this is *response priming* (Klotz & Neumann, 1999; Klotz & Wolff, 1995). In response priming, participants perform a speeded response to a target stimulus that is preceded by a prime stimulus triggering either the same response as the target *(consistent prime)* or the opposite response *(inconsistent prime)*. Typically, consistent primes speed up responses to the target while inconsistent primes slow down responses, and this *priming effect* increases with increasing time interval between prime onset and target onset *(stimulus-onset asynchrony, SOA;* Vorberg, Mattler, Heinecke, Schmidt, & Schwarzbach, 2003). One interesting feature of this paradigm is that response priming can continue to increase with SOA even if the visibility of the prime is decreasing with SOA (Vorberg et al., 2003; Mattler, 2003; Albrecht, Klapötke, & Mattler, 2010), creating a striking double dissociation between motor activation and visual awareness (Schmidt & Vorberg, 2006; Schmidt, 2007; Schmidt et al., 2010).[1]

Response priming effects occur because the prime activates the response assigned to it, inducing a motor conflict if prime and target are inconsistent. This has first been demonstrated in the time course of lateralized readiness potentials in the EEG *(LRPs)*, which start out time-locked to the prime, first develop in the direction specified by the prime, and only later proceed in the direction specified by the actual target (Eimer & Schlaghecken, 1998; Klotz, Heumann, Ansorge, & Neumann, 2007; Leuthold & Kopp, 1998; Vath & Schmidt, 2007; Verleger, Jaśkowski, Aydemir, van der Lubbe, & Groen, 2004). An equally effective way to trace the prime's motor impact over time is the analysis of primed pointing responses (Brenner & Smeets, 2004; Schmidt, 2002) or of response forces in isometric keypress movements (Schmidt, Weber, & Schmidt, 2014). Again, these studies show that inconsistent primes are able to mislead response processes into the wrong direction, such that the initial finger movement is time-locked to the prime, first proceeds in the direction specified by the prime, and only then proceeds in target direction. These effects clearly show that the priming effect increases with prime-target SOA because the prime has progressively more time to direct the response into the correct or incorrect direction, or even to provoke a response error. Consequently, response errors tend to occur only in inconsistent trials at long SOAs.

**2. Feedforward criteria and rapid-chase systems**

The above studies have led to the discovery of a rather stunning invariance: The early time-course of typical response priming effects depends only on properties of the prime, but is completely independent of all properties of the actual target. This invariance has been demonstrated for the discrimination of natural images of animals vs. non-animals, the discrimination of simple geometric objects (both Schmidt & Schmidt, 2009, see Fig. 1 and 2, also see Haberkamp, Schmidt, & Schmidt, 2013; Haberkamp & Schmidt, 2014), and for red and green color stimuli (Schmidt, Niehaus, & Nagel, 2006).

---

[1] See Schmidt, Haberkamp, and Schmidt (2011) for some methodological recommendations on response priming.

Furthermore, the invariance has been demonstrated in pointing movements (Schmidt & Schmidt, 2009, 2010; Schmidt & Seydell, 2008), lateralized readiness potentials (Vath & Schmidt, 2007) and in the time-course of force development in isometric keypress responses (Schmidt, Weber, & Schmidt, 2014). The early invariance of the priming effect is a strong prediction of feedforward models of response priming that are based on the idea that primes and targets trigger sequential feedforward sweeps that carry all the way through to the overt motor response. Such models predict an early phase of the response that is controlled exclusively by the prime, whereas only later segments of the response should be affected by the actual target. In turn, the early invariance of the effect seems difficult to explain without referring to sequential processing of primes and targets.[2]

Schmidt et al. (2006) proposed a *rapid-chase theory of response priming* (RCT) where primes and targets elicit feedforward sweeps that traverse the visuomotor system in strict sequence, without any temporal overlap. Prime and target sweeps are able to directly initiate the motor responses assigned to them, with no need for conscious control (principle of *direct parameter specification,* Neumann, 1990). The prime signal reaches motor areas first, initiating a response and continuing to drive the response on its own until the target signal takes over response control. Priming effects, as well as error rates in inconsistent trials, increase with prime-target SOA because the prime has more time to drive the response on its own when the target is further delayed. Such a feedforward system has to meet the following *feedforward criteria* (Schmidt et al., 2006):

  *A) Initiation criterion:* Prime rather than target signals determine the onset and initial direction of the response.

  *B) Takeover criterion:* Target signals influence the response before it is completed.

  *C) Independence criterion:* Movement kinematics initially depend on prime characteristics only and are independent of all target characteristics.

A stimulus-response system meeting the feedforward criteria is *behaviorally equivalent* to a simple feedforward system and is called a *rapid-chase system*.

---

2 Another prediction of a feedforward system is that priming effects should be fully present in the fastest motor responses, for example, in the fastest deciles of the response time distributions from consistent and inconsistent trials. This has been demonstrated for priming by real and illusory contours (Seydell-Greenwald & Schmidt, 2012), priming by closure of contours (Schmidt & Schmidt, 2014), and priming by occluded contours (Schmidt, Weber, & Schmidt, 2014).

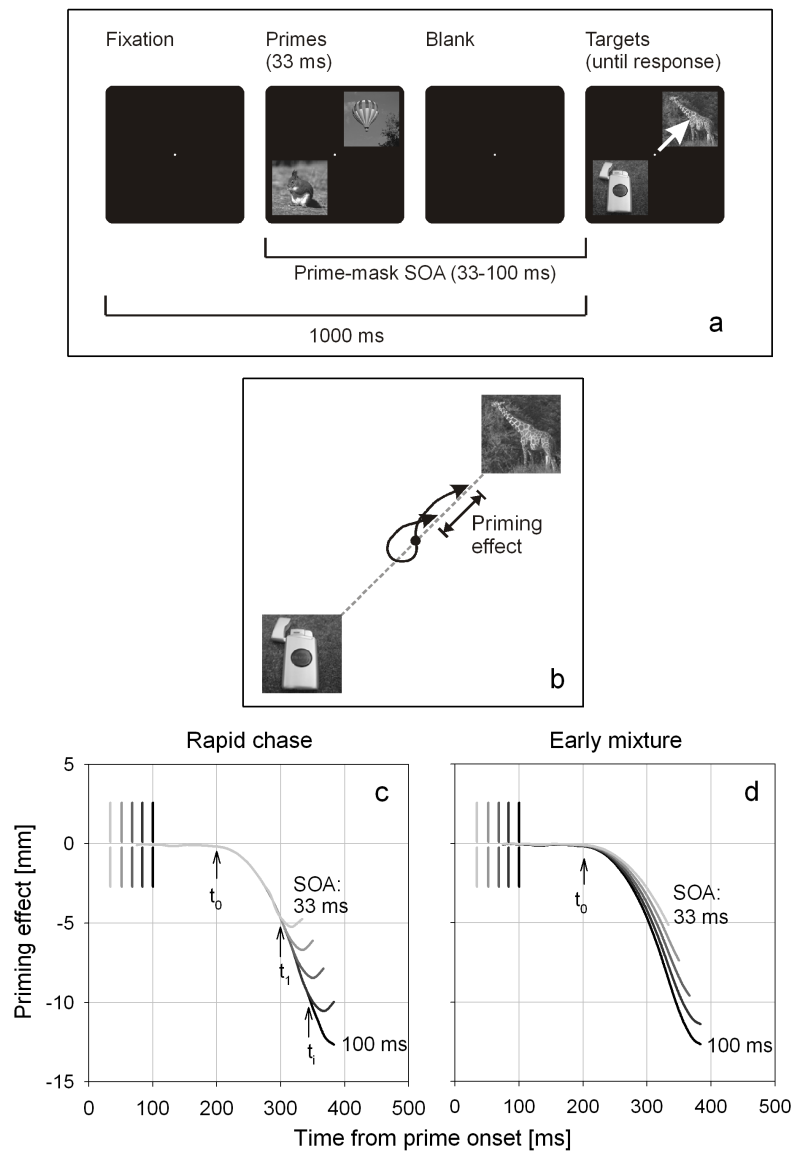

*Fig. 1a: In each trial, two target stimuli (one animal, one object) are presented simultaneously, and the task is to point as quickly as possible from the central position to the animal picture. Targets are preceded by two primes from the same two picture categories at the same locations. Image categories can be spatially switched compared to the targets (consistent trials), or they can agree with the targets (consistent trials), so that the primes trigger a response in the incorrect or correct direction, respectively. b) Spatial priming effects are defined as the spatial difference of the finger positions in inconsistent minus consistent trials. Negative values indicate that the finger position in consistent trials is closer to the target than it is in inconsistent trials at the same point in time. c) Spatial priming effects meeting the feedforward criteria. If response control is strictly sequential, the initial time-course of the pointing response must be invariant when time-locked to prime onset, because there must be a stretch of time where the response is determined exclusively by the prime. Priming trajectories for progressively later prime-target SOAs branch off one by one from an otherwise invariant time-course. Onset of the priming effect is at $t_0$, the branch-off times are at $t_i$. Prime onset is at time 0 so that all trajectories are time-locked to the prime; possible target onsets are indicated by vertical bars. d) Spatial priming trajectories failing the feedforward criteria. The shorter the prime-target SOA, the smaller the impact of the prime on the motor response, and the larger the impact of the target. Therefore, priming trajectories are expected to form a fan-shaped pattern right from the start of the overt response. (From Schmidt & Schmidt, 2009.)*

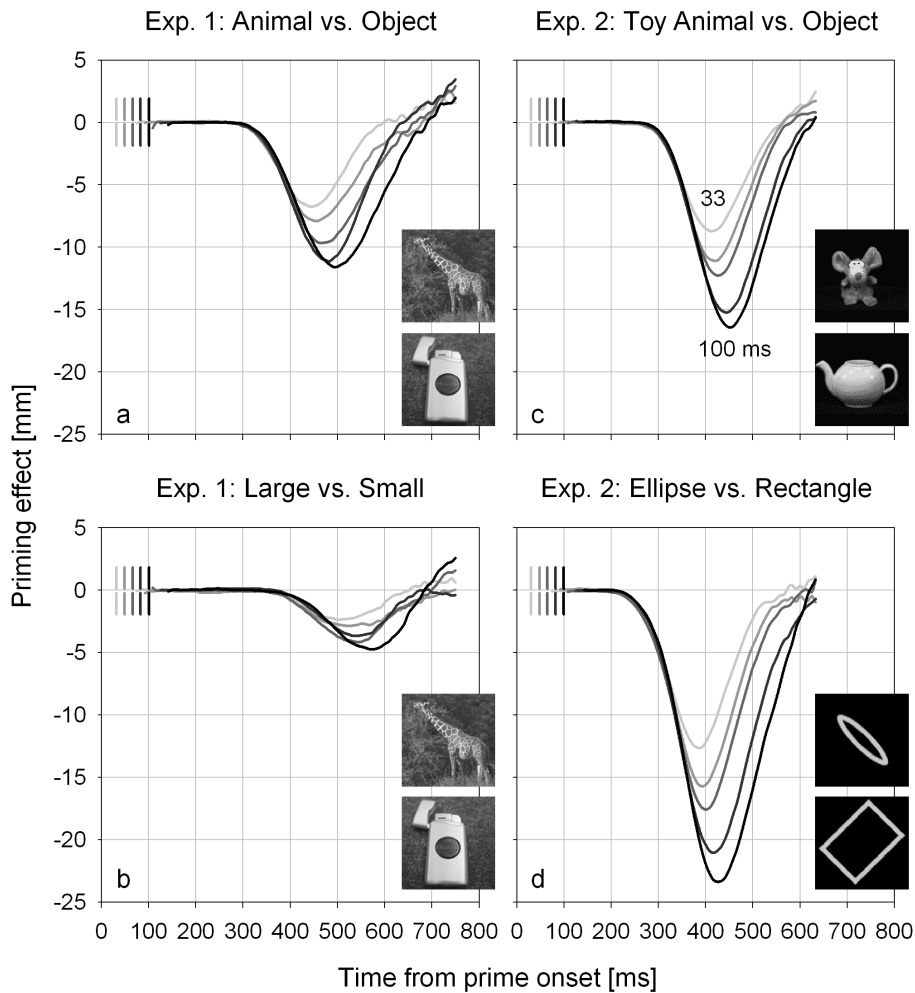

*Fig. 2. Priming trajectories in different experimental tasks (from Schmidt & Schmidt, 2009). a) Natural images of animals vs. objects meet RCT's feedforward criteria. b) So do natural images of toy animals vs. objects (originally in color, taken from a picture set by Geusebroek, Burghouts, & Smeulders, 2005). c) Priming effects break down and cease to meet the feedforward criteria when participants have to judge which of two objects would be larger in real life, a rather awkward, time-consuming categorization. d) Rectangles vs. ellipses can be categorized the fastest, produce the earliest and largest priming effects, and meet the feedforward criteria.*

### 3. Formalizing the feedforward criteria

To formalize the feedforward criteria, we consider measures of response activation that unfold continuously over time so that their time-course can be measured, such as the finger position in a pointing response, a lateralized readiness potential, or the development of response forces in an isometric keypress response.

Consider Fig. 1c. Let $t_\pi$ and $t_\tau$ indicate the onset times of the prime and target, respectively, and SOA = $t_\pi - t_\tau$. We define time $t$ as time-locked to prime onset, $t \equiv t - t_\pi$. Let $\pi$ be a vector of prime properties that may influence the response (such as intensity, duration, color saturation), $\tau$ a vector of target properties, and $i$ the index number of an ordered set of prime-target SOAs, $i = \{1..n\}$, $SOA_i < SOA_j$ for all $i < j$. Let $f_{con,i}(t,\pi,\tau)$ and $f_{inc,i}(t,\pi,\tau)$ be the time-course of response activation in response-consistent and response-inconsistent trials at $SOA_i$, respectively, given the stimulus properties of prime and target.

We define the time course of the priming effect (the *priming trajectory*) as $f_i(t,\pi,\tau) = f_{inc,i}(t,\pi,\tau) - f_{con,i}(t,\pi,\tau)$. Let $|f_i(t,\pi,\tau)|$ describe the *magnitude* of the priming effect (with a sign appropriately defined from the idea of the experiment). Finally, let $t_0$ indicate the onset of the priming effect, and $t_i$, $i = \{1..n-1\}$, the points in time where priming trajectories from individual SOAs branch off from the otherwise invariant time-course.

Rapid-chase theory requires that the sequence of prime and target leads to strictly sequential response control by the two stimuli. We classify the stimulus-response system as a *rapid-chase system* if any pair of empirically measured priming trajectories, $f_i$ and $f_j$, at SOAs $i$ and $j$, $i < j$, satisfies the *feedforward criteria*:

**(1)**    $t \leq t_i$:    $f_i(t,\pi,\tau) = f_j(t,\pi,\tau) = f_0(t,\pi) \neq 0$    *(initial invariance)*
**(2)**    $t > t_i$:    $j > i \rightarrow |f_j(t,\pi,\tau)| > |f_i(t,\pi,\tau)|$    *(ordered branching)*

(That's right: if we formalize them, we only need two lines.) So, for any pair of priming trajectories time-locked to prime onset, both priming effects follow the same invariant initial time course, $f_0$, until the priming trajectory for the shorter SOA branches off while the priming effect for the longer SOA continues to grow (Fig. 1c). Note that $f_0$ depends only on prime but not on target characteristics and must be different from a flatline.

Some corollaries follow from these criteria:

**(3a)**    $t_0 = $ **constant**
The onset of the priming effect is time-locked to the prime and not influenced by properties of the actual target.

**(3b)**    $t_1 < t_2 < ... < t_{n-1}$
Priming trajectories branch off one by one from the common time course in the order of their SOAs. (Note that $t_n$ remains undefined because there is only one trajectory remaining.)

**(3c)**    $t \leq t_i$:    $f_i(t,\pi,\tau) = f_j(t,\pi,\tau) = f_i(t,\pi) = f_j(t,\pi)$
The initial, invariant part of the priming trajectory may depend on properties of the prime but is independent of all target properties.

**(3d)**    $t \leq t_1$:    $f_i(t,\pi,\tau) = f_0(t,\pi)$
In addition, all priming trajectories are identical before the trajectory for the shortest SOA branches off.

**(3e)**    $t > t_i$:    $|f_1(t,\pi,\tau)| \leq |f_2(t,\pi,\tau)| \leq ... \leq |f_n(t,\pi,\tau)|$
After some priming trajectories have branched off from the common time-course, the remaining ones continue to grow.

## 4. Some caveats and considerations

*1) Not all rapid-chase systems are feedforward systems.* This distinction is important because a rapid-chase system is classified on the basis of stimulus input and motor output, not of neuronal processes. Therefore, a rapid-chase system can tolerate some local recurrent processing as long as strictly sequential signals still lead to strictly sequential motor output. As a consequence, the notion of a rapid-chase system is milder than that of a feedforward system: feedforward implies rapid chase, but not vice versa.

*2) Feedforward sweeps by prime and target are not mutually independent.* Rapid-chase processing only proposes that the first sweep must be independent of the second sweep, but not vice versa. Therefore, the criteria demand that initial processing be controlled exclusively by the prime, but not that later processing be controlled exclusively by the target. Independence of the second sweep from the first one is implausible because the second signal passes the exact same neuronal structures that have just processed the first signal, thus meeting the system in a different functional state. As a consequence, the time segment during which priming is invariant may exceed the prime-target SOA. Even though time-locking to the target instead of the prime may reveal interesting relationships, rapid-chase theory does not imply that any part of the priming trajectories be invariant when locked to target onset. Also, it is not predicted that the late parts of consistent trajectories be identical to the late part of inconsistent trajectories.

*3) No invariance predicted in exclusively consistent trials.* It is true that the initial response trajectories should be controlled exclusively by the prime and thus be invariant when locked to prime onset, even when only consistent or only inconsistent trials are considered. The later segments, however, are additionally influenced by the target. Even if prime and target are identical, the target may give a fresh boost to the response process, and targets presented at different times may lead to different trajectories even in exclusively consistent trials.

## Conclusions

Rapid-chase theory offers functional criteria to evaluate whether visuomotor processing is consistent with a simple feedforward process. The criteria are strict and formalized, and they lead to nontrivial predictions about various continuous response measures. Strong evidence for rapid-chase processes as defined here have accumulated for several types of visual stimuli (color, shape, natural objects, contours) as well as several types of response measures (pointing responses, response forces, LRPs, response time distributions). Even though a feedforward sweep in human vision cannot currently be observed directly, our criteria make it possible to bridge the gap between covert neuronal large-scale processing properties and overt motor behavior, perhaps with some pleasant consequences for the understanding of visual perception, visuomotor control, and visual awareness.